\documentclass{pasj01}
\draft

\usepackage{url}
\usepackage{natbib}

\newcommand{\void}[1]{}
\renewcommand{\textcolor}{\void}

\begin{document}
%\SetRunningHead{K.Arimatsu et al.}{OASES project}
%\Received{}%{yyyy/mm/dd}
%\Accepted{}%{yyyy/mm/dd}
%\Published{}%{yyyy/mm/dd}

\title{Organized Autotelescopes for Serendipitous Event Survey (OASES):  design and performance}

%%% begin:list of authors
% Do NOT capitalize all letters in "textsc".
\author{
Ko       \textsc{Arimatsu}\altaffilmark{1}  \\
Kohji \textsc{Tsumura}\altaffilmark{2} 
Kohei \textsc{Ichikawa}\altaffilmark{1,3,4} 
Fumihiko \textsc{Usui}\altaffilmark{5}
Takafumi \textsc{Ootsubo}\altaffilmark{6}  
Takayuki \textsc{Kotani}\altaffilmark{7,1}
Yuki \textsc{Sarugaku}\altaffilmark{8} 
Takehiko \textsc{Wada}\altaffilmark{9} 
Koichi \textsc{Nagase}\altaffilmark{9}  \\
\& \\
Jun-ichi   \textsc{Watanabe}\altaffilmark{1}}
\altaffiltext{1}{
National Astronomical Observatory of Japan, 2-21-1 Osawa, Mitaka, Tokyo 181-8588, Japan}
\email{ko.arimatsu@nao.ac.jp}  
\altaffiltext{2}{Frontier Research Institute for Interdisciplinary Science, Tohoku University, 6-3 Aramaki Aza-Aoba, Aoba-ku, Sendai 980-8578, Japan}
\altaffiltext{3}{Department of Physics and Astronomy, University of Texas at San Antonio, One UTSA Circle, San Antonio, TX 78249, USA}
\altaffiltext{4}{Department of Astronomy, Columbia University, 550 West 120th Street, New York, NY 10027, USA}
\altaffiltext{5}{Center for Planetary Science, Graduate School of Science, Kobe University, 7-1-48, Minatojima-Minamimachi, Chuo-Ku, Kobe 650-0047, Japan}
%Graduate School of Science, Earth and Planetary Sciences, Kobe University, 1-1 rokkodai-cho, nada-ku, Kobe, 657-8501, Japan}
\altaffiltext{6}{Department of Earth Science and Astronomy, The University of Tokyo, 3-8-1 Komaba, Meguro-ku, Tokyo 153-8902, Japan}
\altaffiltext{7}{Astrobiology Center, 2-21-1 Osawa, Mitaka, Tokyo 181-8588, Japan}
\altaffiltext{8}{Kiso Observatory, Institute of Astronomy, Graduate School of Science, The University of Tokyo 10762-30, Mitake, Kiso-machi, Kiso-gun, Nagano 397-0101, Japan}
\altaffiltext{9}{Institute of Space and Aeronautical Science,
Japan Aerospace Exploration Agency, 3-1-1 Yoshinodai, Chuo-ku, Sagamihara, 
Kanagawa 229-8510, Japan}

%% `\KeyWords{}' always has to be placed before `\maketitle'.
 %Do NOT move this preamble from here!

\maketitle

\begin{abstract}
Organized Autotelescopes for Serendipitous Event Survey (OASES) 
is an optical observation project that aims to detect and investigate stellar occultation events 
by kilometer-sized trans-Neptunian objects (TNOs).
%Since the duration of such stellar occultation is smaller than one second,
%and the rate expected to be 
In this project,
multiple low-cost observation systems for wide-field and high-speed photometry
were developed in order to detect rare and short-timescale stellar occultation events.
%consisting of commercial off-the-shelf optics 
%and a CMOS imaging camera monitors large number of stars simultaneously. 
The observation system consists of commercial off-the-shelf  $0.28 \ {\rm m}$  aperture $f/1.58$ optics 
providing a $2\degree.3 \times 1\degree.8$ field of view.
A commercial CMOS camera is coupled to the optics 
to obtain full-frame imaging with a frame rate greater than $10 \ {\rm Hz}$. 
As of September 2016, this project exploits two observation systems, 
which are installed on Miyako Island, Okinawa, Japan.
Recent improvements in CMOS technology in terms of high-speed imaging 
and low readout noise mean that
the observation systems are capable of monitoring 
$\sim 2000$ stars in the Galactic plane simultaneously with magnitudes down to ${\rm V} \sim 13.0$, 
providing $\sim 20\%$ photometric precision in light curves with a sampling cadence of $15.4  \ {\rm Hz}$. 
This number of monitored stars is larger than for any other existing instruments for coordinated occultation surveys.
In addition, a precise time synchronization method needed for simultaneous occultation detection
is developed using faint meteors. 
The two OASES observation systems are executing coordinated monitoring observations
of a dense stellar field in order to detect occultations by kilometer-sized TNOs for the first time.
\end{abstract}

\section{Introduction}
\label{intro}
More than $10^{11}$ kilometer-sized
(hereafter km-sized, with 
\textcolor{red}{radius} $\textcolor{red}{r} \sim 1-10 \ {\rm km}$) 
trans-Neptunian objects (TNOs) are thought to lie in the Kuiper belt and beyond, 
i.e., the scattered disk  \citep{duncan97, volk08}, and the Oort cloud \citep{oort50}. 
The TNOs are thought to be remnants of icy planetesimals  
and thus are of key importance for the study of the early phase of the outer solar system. 
Recent collisional-evolution models by \citet{schlichting13} have shown that the size distribution of the 
present Kuiper-belt objects still contains the signatures of the size distribution of the planetesimals. 
Since the planetesimal size before runaway growth
is one of the major open issues for planet formation theories \citep{chiang10},
the size distributions of km-sized TNOs
would thus give fundamental knowledge about
the accretion processes from the early stages of the solar system.
In addition, determining the present size distribution of the km-sized TNOs
will provide the information on the present-day impact and cratering
rate on the outer planets and their satellites. 
With the cratering rate, 
it will be possible to determine the age of the surface accurately \citep{greenstreet15}. 
Therefore, determining the abundance of the km-sized TNOs is essential 
for understanding the surface evolution of outer solar system bodies, 
especially of bodies in the Pluto system,
whose surface conditions were recently revealed by 
the New Horizons spacecraft \citep{stern15}.

As of 2016, the size distribution of these small-sized TNOs is still unclear.
So far, more than 1500 TNOs have been discovered 
by optical observations in the Kuiper belt and the scattered disk region
\citep[e.g.][]{jewitt93}.
However, these TNOs are larger than \textcolor{red}{10} km \citep[e.g.][]{fraser08}, 
and no km-sized TNOs have been detected so far. 
Direct detection of km-sized TNOs is challenging
because they are extremely faint, with typical optical magnitudes fainter than \textcolor{red}{28}, 
and are thus undetectable even using telescopes with apertures $\sim 10$ m.
Therefore, the size distribution of km-sized TNOs 
still remains an open question.

Instead of direct detection, 
monitoring of stellar occultation events 
is one possible way to discover km-sized TNOs. 
%Since a TNO moves in the sky, 
%it occasionally passes in front of stars and other point-like sources.
%At these moments, the signal intensity from the objects changes due to the occultation.
Since a TNO moves in the sky, 
it occasionally passes in front of stars and other point-like sources, 
changing the signal intensity from these objects by occultation.
Several serendipitous studies of occultation events 
have been performed with ground- and space-based observatories
and several optical studies have found possible occultation events.
An analysis using high-cadence ($40 \ {\rm Hz}$) photometry data of guide stars obtained 
with the guidance sensor 
onboard the Hubble Space Telescope has discovered two possible
occultation events by sub-km-sized Kuiper belt objects \citep{schlichting09,schlichting13}. 
A very recent study by \citet{liu15} found 13 possible occultation events by sub-km-sized TNOs
from time-sequence photometry data for asteroseismological studies obtained with the CoRoT satellite.
While these studies have discovered possible occultation events by sub-km sized TNOs,
there has been no detection of occultations by larger, km-sized TNOs, 
which are expected to be much less frequent events. 

In order to detect stellar occultation events by km-sized TNOs for the first time,
monitoring observations face several challenges.
First, observations must monitor a large number of stars simultaneously 
to detect rare occultation events.
Second, signals from stars must be sampled with a cadence faster than $10 \ {\rm Hz}$, 
because the durations of the occultations are expected to be smaller than a second.
%(see section 2).
Occultation observations thus require detector systems 
capable of achieving photometry with time resolution shorter than 0.1 seconds.
Third, detections of stellar occultation events with ground-based instruments 
must be robust to atmospheric scintillation effects.
Unlike space-based observations, 
ground-based monitoring observations suffer from
atmospheric scintillation that can cause false positive detections.

This paper describes an approach to 
detect and explore TNO occultation events 
using multiple small observation systems.
These multiple systems enable us to observe occultation events 
simultaneously and allow us to achieve detections 
that are robust with respect to atmospheric scintillation. 
The recent development of low-cost commercial CMOS detectors and powerful data storage 
allows us to create mobile observation systems providing wide-field and high-cadence observational data.
As an example of the mobile observation systems, 
we discuss a project under current development for detecting serendipitous astronomical events,  
the Organized Autotelescopes for Serendipitous Event Survey (OASES).
The OASES project plans to monitor
known stars using multiple $30$-cm-scale mobile telescopes with 
low-cost CMOS cameras placed on the Miyako Islands, Okinawa, Japan.
Section 2 presents an overview of TNO occultations, 
reviewing the expected stellar light curve profiles for different sampling cadences.
In Section 3, we describe the concept of the OASES project,
and present the observation systems developed for the monitoring observations.
The performance test observations and current monitoring observations 
together with the photometric performance of the OASES observation systems
are presented in Section 4. 
Using the results of the monitoring observations, 
we have developed a new time calibration method for the 
time synchronization between the observation systems.
The details of the calibration method are described in Section 5.
Finally, we summarize the conclusions in Section 6.

%%%%%%%%%%%%%%%%%%%%%%%%%%%%%%%%%%%%%%%%%%%%%%%%%%
\section{General outline of the TNO occultation events}
\label{section2}
%%%%%%%%%%%%%%%%%%%%%%%%%%%%%%%%%%%%%%%%%%%%%%%%%%
Starlight from an occulted star is diffracted by the edges of TNOs
with their %diameters $D$ less than 10 km  
\textcolor{red}{
radii $r$
}
less than 10 km
because the sizes of the TNOs are comparable to the Fresnel scale, $F$, which is given by

\begin{eqnarray}
F  = \sqrt{\frac{R_h \lambda}{2}}  \sim 1.3 \ {\rm km},
\end{eqnarray}
where $R_h = 40 \ {\rm au}$ is the semi major axis of the TNO
($R_h$ is approximated as the distance between the observer and the TNO in this equation), 
and $\lambda = 500 \ {\rm nm}$ is the wavelength of the observations.

\begin{figure}[!t]
\begin{center}
 \includegraphics[scale=0.51, angle= 270]{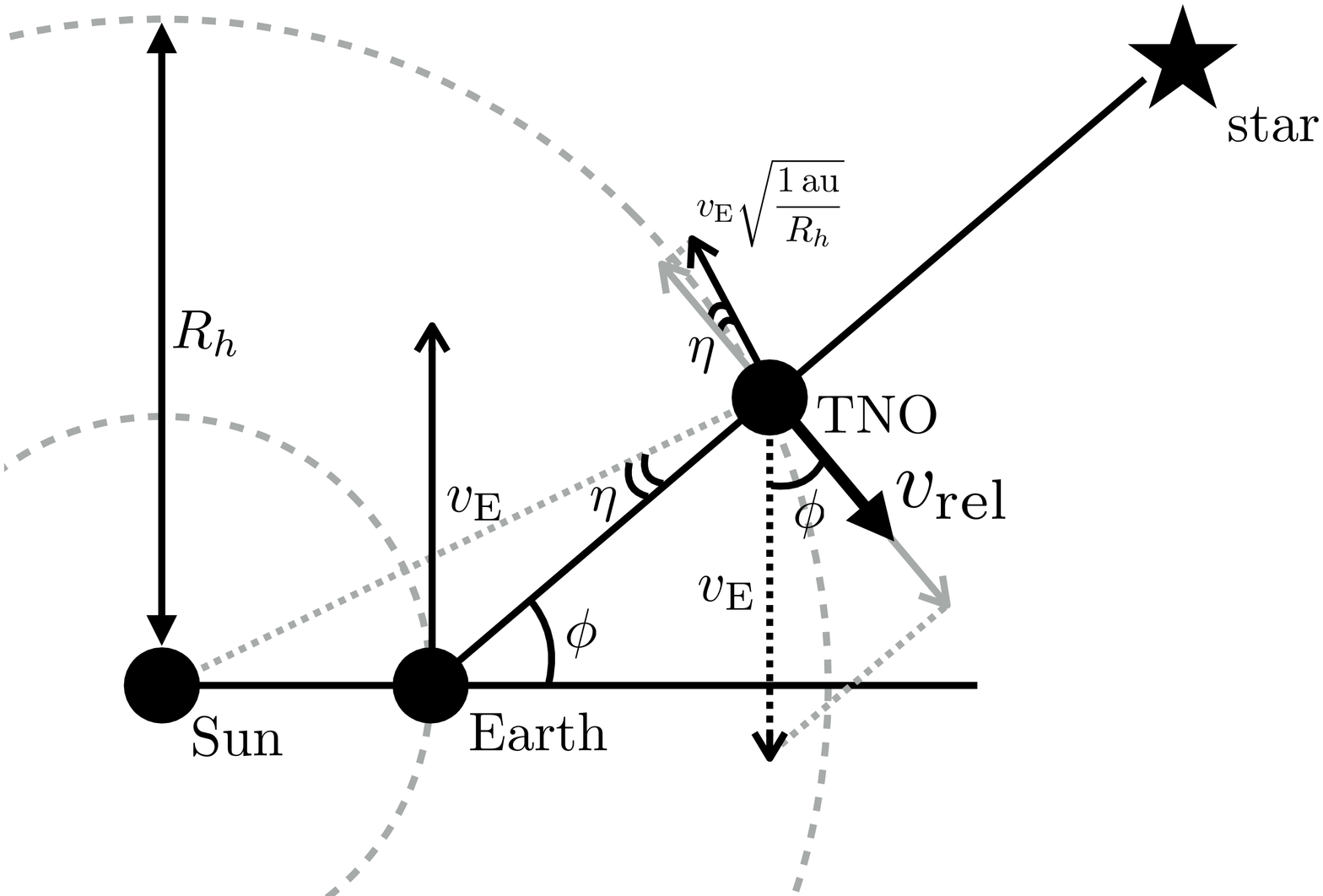}
   \caption{Schematic view of  the orbital geometry used to derive the relative velocity of a TNO $v_{\rm rel}$. 
   The TNO is assumed to be rotating on a circular ecliptic orbit with a radius of $R_h$.}
   \label{fig01}
 \end{center}
\end{figure}

As described in detail by \citet{nihei07}, the observed light curve for an occultation event is 
determined by the size, shape, and distance of the TNO, the angular size of the star, 
the impact parameter between the star and the TNO, 
the wavelength range of the observations, and the relative velocity between the observer and the TNO.
According to the simulations by \citet{nihei07},
the duration of the event is determined by the length of the occulter passing in front of the star
and the relative velocity.
The occulter length $L$ passing in front of a star is described by
\begin{eqnarray}
L = \sqrt{W^2 - (2b)^2},
\end{eqnarray}
where $b$ is the impact parameter and $W$ is the diameter of the occulter, 
which is determined by the %diameter 
\textcolor{red}{
radius
}
of the TNO and the diffraction effects.
Assuming a spherical TNO with a %diameter $D$,
\textcolor{red}{
radius $r$,
}
$W$ is approximated by \citet{nihei07} as
\begin{eqnarray}
%W \sim \bigg[ (2\sqrt{3}F)^{3/2} +  D^{3/2} \bigg]^{2/3},
W \sim \bigg[ (2\sqrt{3}F)^{3/2} +  (\textcolor{red}{2 r})^{3/2} \bigg]^{2/3},
\end{eqnarray}
which corresponds to the diameter of the first Airy ring.
We should note that $W$ does not become 
smaller than $2\sqrt{3}F$, corresponding to $\sim 5\ {\rm km}$ for a TNO 
located in the Kuiper belt ($R_h = 40$ au).
On the other hand,
assuming a TNO on a circular ecliptic orbit with a radius of $R_h$
(see figure~\ref{fig01}),
 the velocity of the TNO relative to the Earth, $v_{\rm rel}$, is given by
\begin{eqnarray}
v_{\rm rel} = v_{\rm E} \cos{\phi} - v_{\rm E} \bigg( \frac{1 {\rm au}}{R_h} \bigg)^{1/2} \cos{\eta}
 = v_{\rm E} \bigg[ \cos{\phi} - \bigg( \frac{1 {\rm au}}{R_h} \bigg)^{1/2} \bigg(1 -  \frac{1 {\rm au}^2}{R_h^2} \sin^2{\phi} \bigg)^{1/2} \bigg],
\end{eqnarray}
where $v_{\rm E}$ is the orbital velocity of the Earth ($29.8 \ {\rm km\ s^{-1}}$) 
and $\phi$ and $\eta$ are the TNO--Earth--opposition and the Sun--TNO--Earth angles in the ecliptic plane, respectively (see figure~\ref{fig01}).
Since the orbital velocity of the TNO is less than $5 \ {\rm km \ s^{-1}}$ and is much smaller than $v_{\rm E}$,
$v_{\rm rel}$ mostly depends on $\phi$.
The duration of the occultation event, $\tau$, is given by 
%\tau  = \frac{W}{v_{\rm rel}} = \bigg( \frac{1}{v_{\rm E}} \bigg) \bigg[ (2\sqrt{3}F)^{3/2} +  D^{3/2} \bigg]^{2/3} \bigg[ \cos{\phi} - \bigg( \frac{1 {\rm au}}{d} \bigg)^{1/2} \bigg(1 -  \frac{1 {\rm au}^2}{d^2} \sin^2{\phi} \bigg)^{1/2} \bigg]^{-1}. 
\begin{eqnarray}
\tau  & = &  \frac{W}{v_{\rm rel}} \\
& = & \bigg( \frac{1}{v_{\rm E}} \bigg) \bigg[ \bigg[ (2\sqrt{3}F)^{3/2} +  \textcolor{red}{(2 r)^{3/2}} \bigg]^{4/3} + (2b)^2 \bigg]^{1/2} \bigg[ \cos{\phi} - \bigg( \frac{1 {\rm au}}{R_h} \bigg)^{1/2} \bigg(1 -  \frac{1 {\rm au}^2}{R_h^2} \sin^2{\phi} \bigg)^{1/2} \bigg]^{-1}. 
\end{eqnarray}
%& = & \bigg( \frac{1}{v_{\rm E}} \bigg) \bigg[ \bigg[ (2\sqrt{3}F)^{3/2} +  D^{3/2} \bigg]^{4/3} + (2b)^2 \bigg]^{1/2} \bigg[ \cos{\phi} - \bigg( \frac{1 {\rm au}}{R_h} \bigg)^{1/2} \bigg(1 -  \frac{1 {\rm au}^2}{R_h^2} \sin^2{\phi} \bigg)^{1/2} \bigg]^{-1}. 
%$\tau$ is thus determined by the 
If a TNO located at opposition ($\phi = 0$) occults a star with $b=0$,  the duration of the occultation event $\tau$ is 
\begin{eqnarray}
%\tau  \simeq 0.2 \bigg[ 1 +  \bigg( \frac{D\ [{\rm km}]}{5} \bigg)^{3/2} \bigg]^{2/3}  \ [{\rm sec}].
\tau  \simeq 0.2 \bigg[ 1 +  \bigg(\textcolor{red}{ \frac{r\ [{\rm km}]}{2.5}} \bigg)^{3/2} \bigg]^{2/3}  \ [{\rm sec}].
\end{eqnarray}
$\tau$ is thus determined by the size of the TNO 
and is expected to be about \textcolor{red}{0.2--0.9} seconds for a km-sized TNO (with \textcolor{red}{$r=1-10 \ {\rm km}$}). 
Therefore, observations with time resolution better than 10 Hz are required
for robust detection of these occultation events.
%In fact, the finite sizes of the stars cannot be ignored for the occultation lightcurves.
%Diffraction patterns are convolved and  are broaden by the stellar disks. 

\begin{figure}[!t]
\begin{center}
 \includegraphics[scale=0.82]{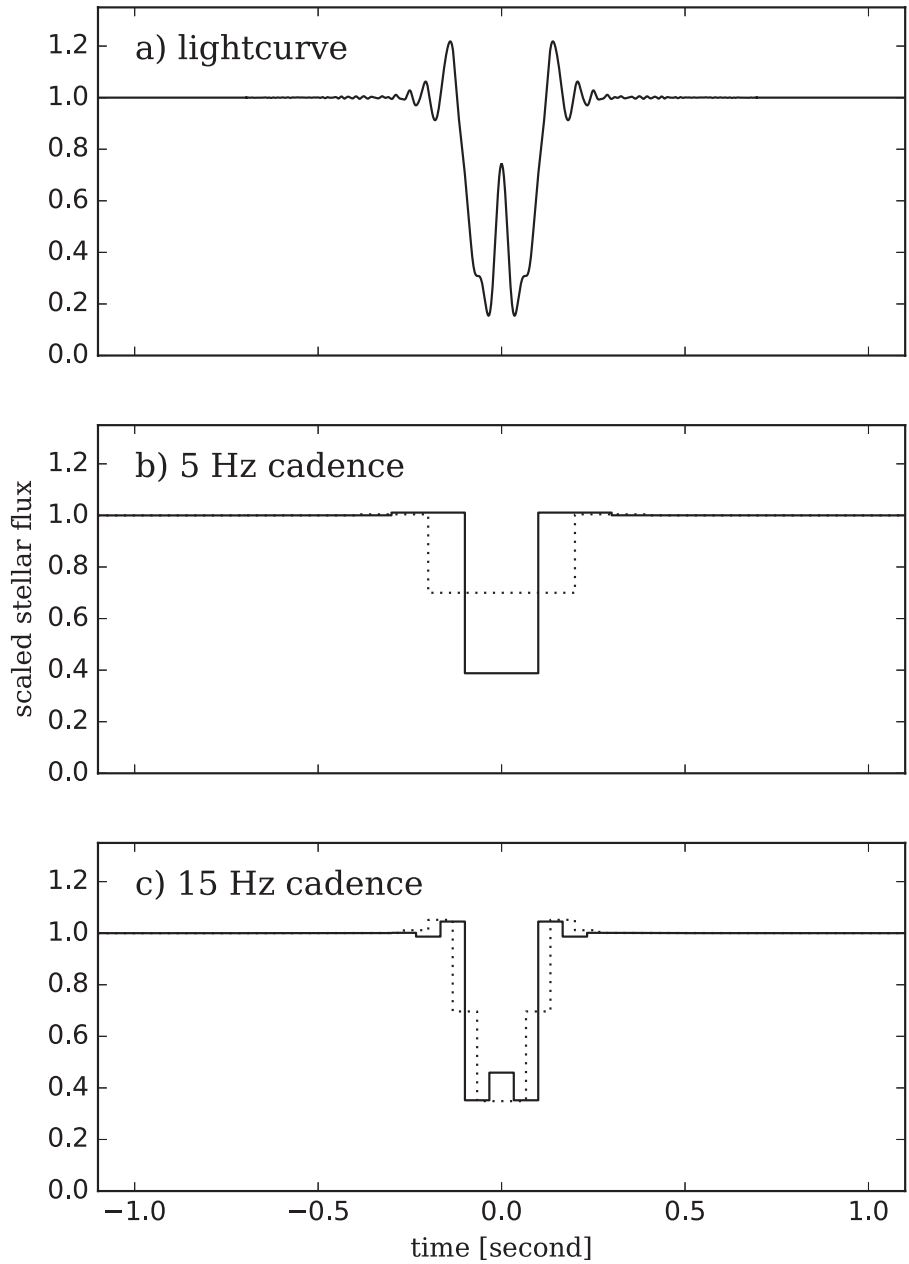}
   \caption{(a): Simulated light curve of an occultation event for a \textcolor{red}{$r = 1.5$}  km sized TNO located at $R_h = 40$ au, 
   and traversed at impact parameter $b=0$. The occultation is assumed to occur at $\phi = 30\degree$, 
   and is observed at a wavelength of $500 \ {\rm nm}$.
    The background object is assumed to be a $V= 13$ mag F6V star.	
   The finite size of the star is taken into consideration.
   (b) and (c): The same light curve  as (a), but sampled at a cadence of (b) 5\ Hz and (c) 15\ Hz.
   Solid and dotted lines in panels (b) and (c) show light curves of the occultation 
   sampled with no time offset and with a time offset of $1/10$ and $1/30\ {\rm seconds}$, respectively.}
   \label{fig1}
 \end{center}
\end{figure}

Figure~\ref{fig1}a shows the expected light curve of a $V= 13.0$ mag F6V star
occulted by a \textcolor{red}{$r = 1.5$} km sized TNO located at 40 au and traversed at an impact parameter $b=0$.
The light curve shows clear diffraction features; 
the local peak is seen at the occultation center, 
and the intensity is enhanced at 
the edges of the shadow corresponding to the first Airy ring. 
Figure~\ref{fig1}b presents the same light curve as figure~\ref{fig1}a but sampled at 5 Hz; 
the same sampling cadence as that of the previous 
ground-based coordinated occultation observations,
Taiwanese American Occultation Survey (TAOS, \citet{lehner09}).
%Since the occultation events occurs short timescale, 
Since the time interval of the occultation is
only $\sim 0.2 \ {\rm sec}$,
%the higher than 5 Hz.
%However, even if we perform sampling cadences of $\sim 5$ Hz, 
%one can detect 
the outlier of the light curve due to the occultation event 
is only detected in one or two bins.
%Therefore the time resolution is not sufficient to perform a robust detection 
%of the occultation events.
%Furthermore, when we perform observations with a sampling cadence of $\sim 5$ Hz, 
Therefore we cannot obtain the information on the duration of the occultations, 
which provides a constraint on the size information of the occulters.
In order to gain a clear constraint of the size of the TNOs, 
we have to obtain the stellar light curves with time resolution 
comparable to, or higher than, $\sim 10$ Hz (figure~\ref{fig1}c).
In fact, ongoing and proposed TNO occultation observation projects such as 
CHIMERA \citep{harding16},
TAOS II \citep{lehner14},  
and the present OASES projects  
plan to monitor stars with sampling cadences of $15-30$ Hz.

As already noted in section~\ref{intro}, stellar occultation by km-sized TNOs is expected to be very rare. 
According to the results of the previous TAOS observations \citep{bianco10, zhang13}), 
the event rate of the occultation is expected to be less than $10^{-2}\ {\rm yr^{-1}}$ per star, 
much smaller than that of the sub-km sized TNO occultations \citep{schlichting13}. 
Therefore serendipitous surveys should monitor more than hundreds of stars simultaneously,
and thus require a wide field.
Furthermore, since stellar occultations by unknown TNOs are short-timescale and non-repeatable phenomena,
the atmospheric scintillation and other noise can cause false detection.
In order to avoid false detection due to noise,  
simultaneous detection by multiple instruments is essential for stellar occultation observations.
%Furthermore, detections of the stellar occultations with multiple separate instruments 
%will help us to obtain the properties of the TNOs.
Increasing the number of observation instruments 
will offer more robust detection of stellar occultation events
because the probability of the false-positive detection significantly decreases with
the number of independent observation data obtained simultaneously \citep{lehner10}.
One of the challenges for the installations of the multiple instruments 
is to realize their low-cost production;
the total cost of the observation project increases almost linearly with the number of instruments.
Therefore the development of multiple low-cost instruments has a key role in
the achievement of coordinated monitoring observations.

%%%%%%%%%%%%%%%%%%%%%%%%%%%%%%%%%%%%%%%%%%%%%%%%%%
\section{Outline of the OASES observation system}
\label{sec:obssystem}
%%%%%%%%%%%%%%%%%%%%%%%%%%%%%%%%%%%%%%%%%%%%%%%%%%

The OASES project uses multiple observation systems, 
each containing a commercial monochromatic CMOS camera 
and an optical tube for astronomical photographs. 
This section describes the observation system 
and the current observation site of the OASES project.
Hardware specifications 
%and imaging and photometric performances 
of the current OASES observation system
are summarized in table~\ref{tab31}.
%The photometric performances are estimated based on the results of the ongoing 
%monitoring observations, which are described in the following section. 
The principal advantage of the system is its cost effectiveness.
Since commercial off-the-shelf optics and CMOS cameras are adopted,
an installation of a single OASES observation system costs about 16000 US \$. %,
%roughly an order of magnitude smaller than the previous TAOS project \citep{weiss98}.
This advantage makes it possible to 
install a large number of multiple observation systems with limited financial resources.
As of September 2016, the OASES project exploits two observation systems (OASES-01 and OASES-02)
and plans to increase the total number of systems in the near future.

\begin{table}
  \caption{Specifications of the OASES prototype and observation systems}\label{tab31}
  \begin{center}
    \begin{tabular}{lcc}
      \hline 
       %\hline
Observation system                                                          & prototype system & observation system \\
%   \hline 
%   \multicolumn{3}{c}{Hardware}  \\
   \hline 
 Number of systems                                                                     & 1                  & 2 \\
 Optics                                                                      & \multicolumn{2}{c}{Celestron, LLC. Rowe-Ackermann Schmidt astrograph} \\
 \ \ Aperture  [mm]                                                                  & \multicolumn{2}{c}{279}                      \\
 \ \ Focal Ratio                                                                       &  f/2.22   &    f/1.58                  \\
\ \ Camera                                                                  &   Point Grey Research, Inc. &  ZWO Co., Ltd.    \\
                                                                                &   {\it grashopper 3} GS3-U3-23S6M-C    &         ASI1600MM-Cool  \\
 \  \ Number of effective pixel                                      & $1920\times1200$        &     $4656\times3520$    \\
 \  \ \textcolor{red}{  pixel size  [${\rm \mu m}$] }         &     \textcolor{red}{5.86}       &   \textcolor{red}{3.8}  \\
 \  \  Angular pixel scale [arcsec ${\rm pixel}^{-1}$]    & 1.96                                &   1.79 (3.59 :$2\times 2$ binned mode)            \\
\  \ Field of view [${\rm degree^{2}}$]               & 0.68                               &    \textcolor{red}{4.05}     \\
\  \ Readout noise [$e^-$]                                           & 6.8                                    &     2.9   \\
\  \ Quantum efficiency [\% at 525 nm]                       & 76                                     &     $\sim 60$   \\
Mount                                                                         &  \multicolumn{2}{c}{Takahashi Seisakusho Ltd. EM-200 {\it Temma-2 Jr.} German equatorial mount} \\ 
 \hline 
    \end{tabular}
  \end{center}
\end{table}

\subsection{Camera component}

\begin{figure}[!t]
\begin{center}
   \includegraphics[scale=0.85]{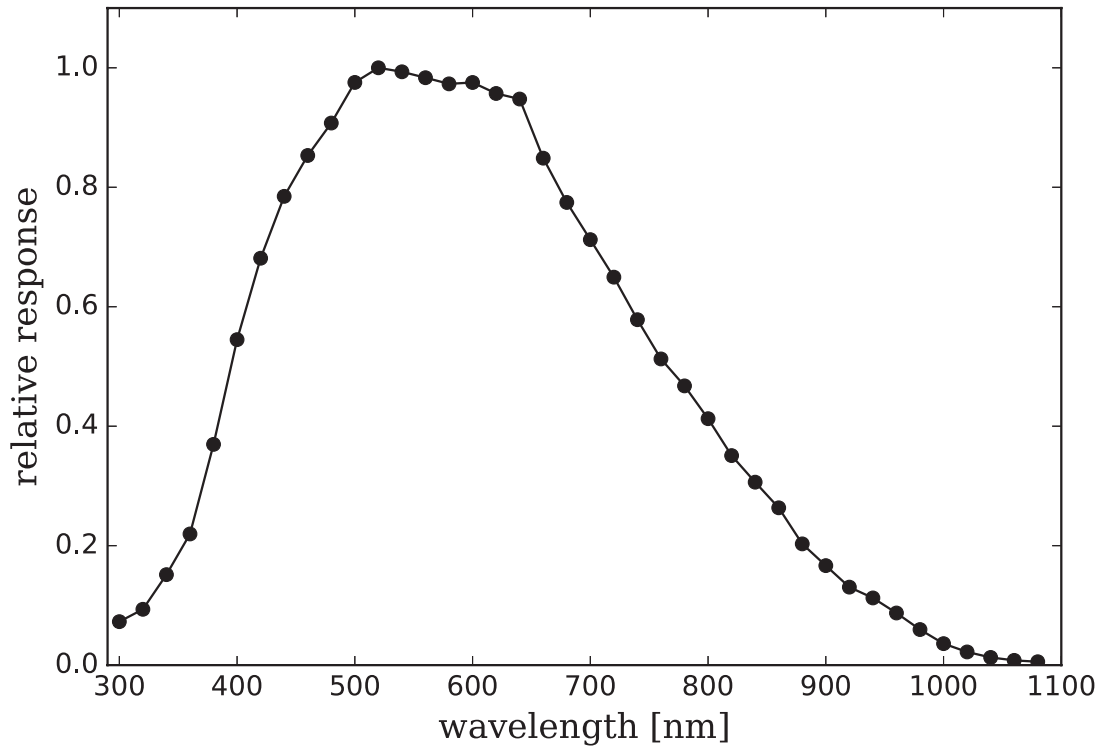}
   %\caption{Quantum efficiency of the SONY IMX174 (solid curve) and IMX250 (dashed curve) CMOS detectors.}
   \caption{The measured relative spectral response of the ZWO Co., Ltd. ASI1600MM-Cool CMOS camera.
   The response is normalized at the peak.}
   \label{fig32}
 \end{center}
\end{figure}

The OASES observation system 
employs a ZWO Co., Ltd. ASI1600MM-Cool monochromatic CMOS camera.
This camera has a MN34230 
front-illuminated CMOS sensor manufactured by the Panasonic Corporation.
The CMOS sensor consists of $4656 \times 3520$ square pixels 
with a pixel pitch of $3.8 \ {\rm \mu m}$, 
offering a $17.7 \ {\rm mm} \times 13.4 \ {\rm mm}$ effective area.
%The quantum efficiencies of the IMX174 
%and 
\textcolor{red}{
According to the manufacturer of the CMOS sensor
(\url{https://industrial.panasonic.com/content/data/SC/PDF/news2013/jp/IS00006AJ.pdf}), 
each pixel of the CMOS sensor 
includes a microlens to increase the effective photon collecting area.
}
Since the spectral response of the CMOS camera 
%for a sufficient wavelength range
is not supplied by the manufacturer in detail, 
it was measured 
using a quantum efficiency (QE) measurement bench \citep{kamata04}.
The relative spectral response of 
the camera detector normalized at the peak (at $510 \ {\rm nm}$) 
is shown in figure~\ref{fig32}.
%normalized at the peak (at $510 \ {\rm nm}$).
According to the manufacturer 
(\url{https://astronomy-imaging-camera.com/products/usb-3-0/asi1600mm-cool/}),
the absolute peak QE is $\sim 60\%$.
The camera provides an efficiency of over 50\% of the peak value
(corresponding to $\sim 30\%$ absolute QE) 
over the wavelength range of  $400-750 \ {\rm nm}$.
\textcolor{red}{
%The low QE relative to a scientific grade CCD sensors 
The lower QE relative to those of the scientific-grade CCD sensors 
(typically $> 50\%$ absolute QE over the wavelength range of  $400-800 \ {\rm nm}$) 
is most likely due to a smaller photon collecting area 
limited by readout electronics on the front-illuminated CMOS sensor.
}
The ASI1600MM-Cool CMOS camera 
has a two-stage Peltier cooler 
to maintain the CMOS sensor with an operating temperature 
at $5-10 \ {}^\circ\mathrm{C}$.

\subsection{Optical component}
The optical component of the OASES observation system 
is a Celestron, LLC. Rowe-Ackermann Schmidt Astrograph (RASA).
The RASA is a prime-focus astrograph
with an effective aperture diameter of  279 mm, 
and focal ratio $f/2.22$.
%The detailed optical design of the astrograph is proprietary.
%However, according to the published sources, 
%the RASA has a primary mirror, and a prime-focus corrector 
%consisting of five lens arranged in four groups. 
The primary advantage of the RASA is that 
it provides a wide optical field of view (FOV) with a very fast focal ratio.
%According to a published source from 
A Celestron, LLC. publication \citep{berry16} 
lists the designed image circle diameter of the astrograph as $\phi = 43.3 \ {\rm mm}$,
which corresponds to a FOV of $4.00\degree$ across 
($12.6 \ {\rm degree^2}$ area).

Although the RASA provides a wide field with the fast focal ratio,
%the RASA provides a very fast focal length,
%the optically flat field is too large for the CMOS sensor 
it is too large for the CMOS sensor 
of the OASES camera component. 
Taking the focal length of the RASA ($620 {\rm mm}$) 
and the effective area of the ASI1600MM-Cool CMOS camera 
($17.7 \times 13.4$ mm) into account,
the FOV covered by the CMOS sensor 
is $1.64\degree \times 1.24\degree$ ($2.02 \ {\rm degree^2}$ area),
which is much smaller than 
the optically flat FOV provided by the RASA optics ($12.6 \ {\rm degree^2}$ area).
In order to increase the instrumental FOV
and to improve the efficiency of monitoring observations,
we insert a focal reducer between the RASA and the CMOS camera.
The Metabones Speed Booster SPEF-M43-BT4
is used as a focal reducer. 
Since the variable ratio of the focal reducer is 0.71, 
the effective focal length is 440 mm (focal ratio $= f/1.58$).
The Speed Booster offers a corrected focal plane
covering the effective area of the CMOS sensor,
and increases the FOV of the observation system to
$2.32\degree \times 1.75\degree$ ($4.05 \ {\rm degree^2}$ area).
However, we found that 
the illumination of the system drops by approximately $60\%$
from the image center to the edge.
We thus generally use the central $\sim 1\degree$-radius field,
over which the illumination drops by only $\sim 20\%$.

In the present observations, no filter wheel is used with the observation system.
%except for a holder that will accommodate 
%a dark shutter manually inserted in front of the camera.
However, we plan to install
%to leave a space for installing
%of 
an additional filter holder in front of the focal reducer in future observations.

\begin{figure}[!t]
\begin{center}
   \includegraphics[scale=0.65]{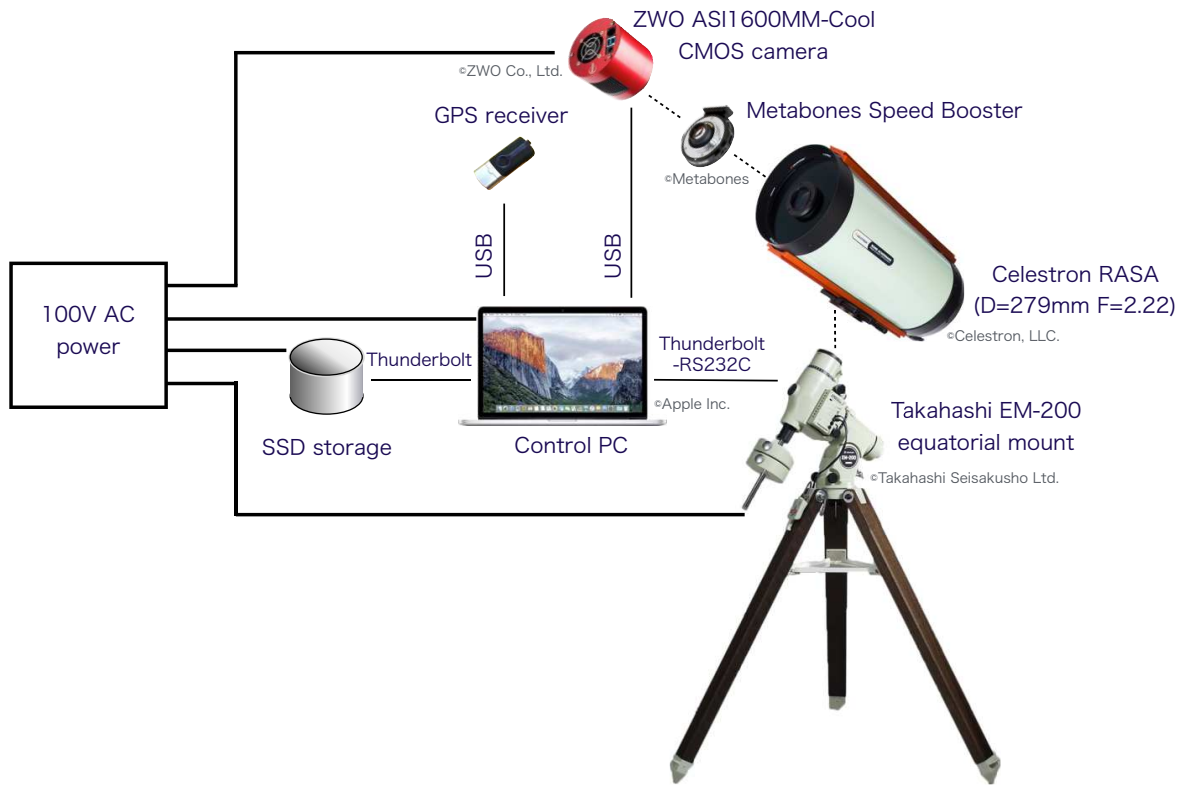}
   \caption{A functional overview of the OASES observation system. 
   The ZWO Co., Ltd. ASI1600MM-Cool CMOS camera is mechanically connected to the Celestron, LLC. RASA via the Metabones Speed Booster focal reducer.
   The CMOS camera and the mount are connected to the same control PC via USB and 
   Thunderbolt-RS232C connections using Thunderbolt to USB and USB to RS232C adapters, respectively.
   The 100 V AC power is supplied to all the electrical devices in the system.}
   \label{fig33}
 \end{center}
\end{figure}

\subsection{Functional overview and control software}
A functional overview of the OASES observation system is presented in
figure~\ref{fig33}.
The individual observation system is mounted on a Takahashi Seisakusho Ltd. EM-200 {\it Temma-2 Jr.} 
German equatorial mount.
The CMOS camera and the equatorial mount are controlled by the same control PC.
\textcolor{red}{
In the present observations, tracking is performed in open loop mode
(without feedback from any guiding facility).
According to the results of the observations, 
the typical root-mean-square tracking accuracy of the mount is $\sim 14\arcsec $ for 30 minutes.
}
The CMOS camera is set up with
{\it Firecapture} version 2.4, an image capture program for USB-connected cameras 
(see \url{http://firecapture.wonderplanets.de}).
The user can control the camera parameters,
and obtain imaging data with user-defined exposure times and 
number-of-frames 
by running command scripts on the software.
The images obtained are recorded on a SSD storage,
which is connected to the control PC via a Thunderbolt connection.
The equatorial mount is 
controlled via a proprietary program, 
{\it StellaNavigator} version 10
(see \url{https://www.astroarts.co.jp/products/software.shtml}).
The user points the observation system to the desired position using the program.

A GPS receiver connected via USB is used as a source for time synchronization of the control PC.
%a GPS receiver connected via USB connection is used in the system.
The internal clock of the control PC is synchronized 
with the GPS time information received from the GPS receiver
using {\it Satk} version 3.4 time adjustment software
(see \url{http://sendaiuchukan.jp/data/occult/gpsradio/satk.html}).
The {\it Firecapture} software writes timestamps for individual frames in the 
header information of the imaging data (SER file).
%The time of the PC's OS is maintained by a GPS receiver connected via USB connection.
The equatorial mount, the CMOS camera, the control PC and the SSD storage 
are powered by the same 100 V AC power supply.

\subsection{Observation site}

\begin{figure}[!t]
\begin{center}
   \includegraphics[scale=0.95]{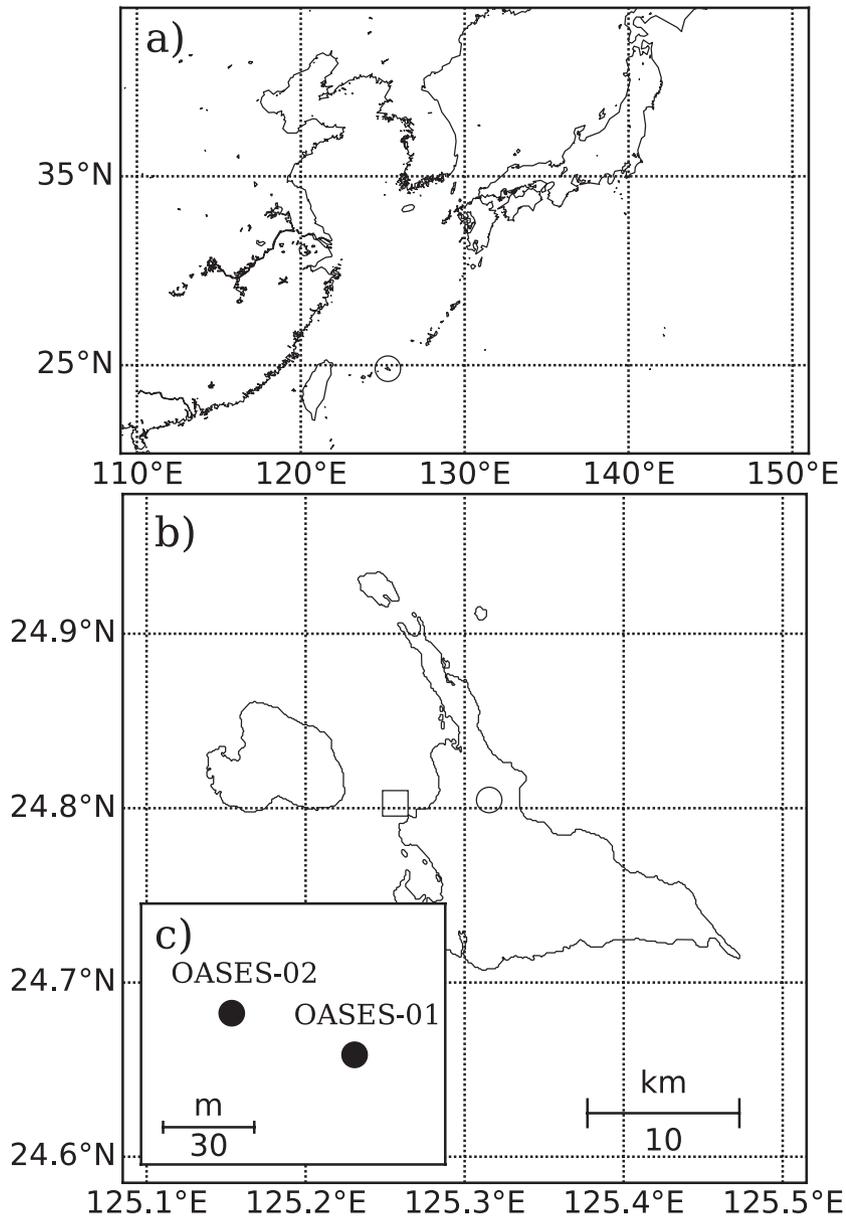}
   \caption{
   (a) Map of East Asia with location of Miyako Island (open circle).
   (b) Map of Miyako Island overlaid with the OASES monitoring site
   (Miyako open-air school: open circle), 
   where the monitoring observations were performed in 2016, 
   and the site of the performance test observations in 2015
   ({\it Toriba Kaihin Koen} park: open square).
   Since the {\it Toriba Kaihin Koen} park was built over reclaimed shoreline, 
   it appears to lie over the sea. 
   In (a) and (b), the coastline data 
   provided with the Basemap Matplotlib Toolkit version 1.0.8
   (\url{http://matplotlib.org/basemap/index.html}) are used.
   (c) The layout of the two OASES observation systems (OASES-01 and OASES-02) at Miyako open-air school.}
   \label{fig34}
 \end{center}
\end{figure}

The two OASES observation systems are currently installed at an
observation site in Miyako Island (figure~\ref{fig34}a and \ref{fig34}b), Miyakojima-shi, Okinawa prefecture, Japan (Miyako open-air school 
%({\it Miyako Shonen Shizen no Ie}),
\textcolor{red}{
({\it Miyako Seishonen no Ie}),
}
 latitude: $24\degree$ 48' 17"N, longitude: $125\degree$ 18' 55"E, altitude: $33 \ {\rm m}$, 
marked as an open circle in figure~\ref{fig34}b).
Since Miyako Island is at $24-25\degree$ N
($\sim 10\degree$ farther south than the major islands of Japan), 
it is one of the most appropriate sites in Japan for the observation of objects near the ecliptic.
In addition, according to the results of the present monitoring observations,
the ratio of fine weather is $\sim 40$\% for the period from late June to early September 
during which the selected observation field (see subsection~\ref{subsec:obs}) is visible. 
This fine weather ratio is higher than at other major observation sites in Japan.
Furthermore, in this period, the island is almost free from strong jet streams, 
which often cause turbulence and may increase atmospheric scintillation effects.

%%%%%%%%%%%%%%%%%%%%%%%%%%%%%%%%%%%%%%%%%%%%%%%%%%%%
\section{OASES observations}
%%%%%%%%%%%%%%%%%%%%%%%%%%%%%%%%%%%%%%%%%%%%%%%%%%%%

\begin{figure}[!pt]
\begin{center}
\includegraphics[scale=0.45]{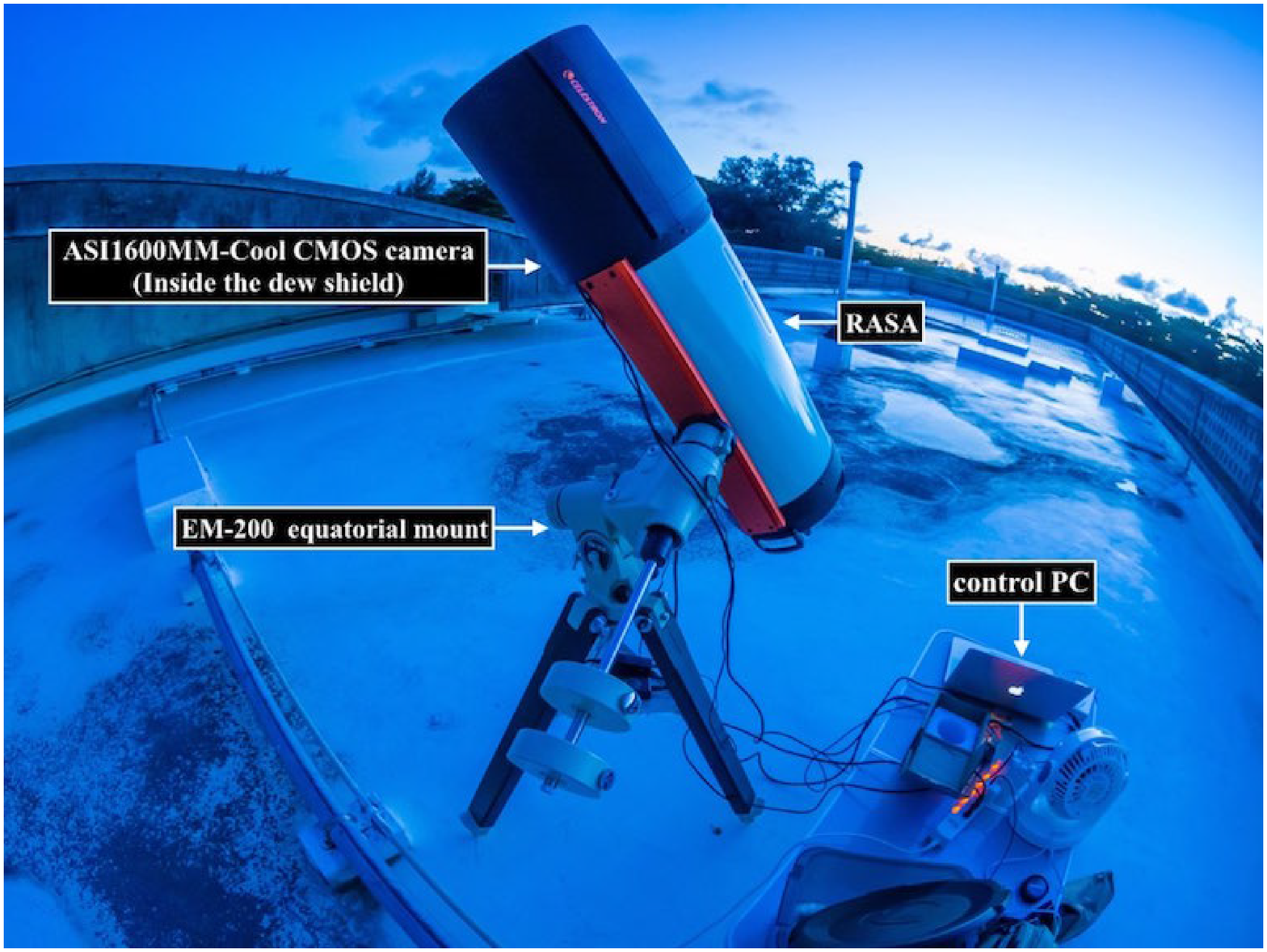}
\caption{The OASES observation system (OASES-02) 
installed at Miyako open-air school, 
Miyakojima-shi, Okinawa.
The system, which consists of a Celestron, LLC. RASA optical tube 
attached to a ASI1600MM-Cool CMOS camera 
(hidden by a dew shield in this image), 
is mounted on a Takahashi Seisakusho Ltd. EM-200 German equatorial mount .}
\label{fig41}
\end{center}
\end{figure}

\subsection{Performance test observations}
\label{subsec:test}
OASES performance test observations were carried out in 2015 
with a single prototype system.
Specifications of the prototype are presented in table~\ref{tab31}.
The prototype was designed with the same optical components as the current system.
However, it is coupled with a small-format CMOS sensor (SONY Corporation IMX174) without a focal reducer
and provides images with a narrower field-of-view ($1\degree .05 \times 0\degree .65$).
The prototype system was placed at "{\it Toriba Kaihin Koen}" park on Miyako Island
(latitude: $24\degree$ 48' 10"N, longitude: $125\degree$ 15' 23"E, altitude: $1 \ {\rm m}$, 
marked as an open square in figure~\ref{fig34}b),
and observed dense stellar fields as proof-of-concept observations.
%We also observed  bright stars with the prototype
%to check the effect of atmospheric scintillation noise on short-timescale photometry.

\subsection{Monitoring observations}
\label{subsec:obs}
The first scientific monitoring observations for stellar occultation began on 25 June 2016 
with the two OASES observation systems.
%This observation systems were placed at "{\it Toriba Kaihin Koen}" park 
%(Lat: $24^\circ$ 48' 10"N, long: $125^\circ$ 15' 23"E, altitude: 16 \, m; 
%see the double-circle in Figure~\ref{fig31}).
The two systems were placed in 
different positions on the rooftop of the Miyako open-air school %({\it Miyako Shonen Shizen no Ie}; 
\textcolor{red}{
({\it Miyako Seishonen no Ie};
}
marked as an open circle in figure~\ref{fig34}b, see also figure~\ref{fig41}).
The two systems were $39 \, {\rm m}$ apart (see Figure~\ref{fig34}c).
%The separation between the two systems was $39 \ {\rm m}$ (see figure~\ref{fig34}c).
%which is sufficient to reduce false detections 
%due to the extreme scintillation noise.
We carried out monitoring for 40 hours in total,
over the period from 25 June to 10 September JST.
%We had carried out monitoring observations for 30 hours in total,
%spread out between 25 June and 10 September JST.

For the monitoring observations, 
we selected a target sky region with the following criteria.
%Since the chances of the occultation increases with the number of stars observed simultaneously, 
Since the chance of occultation increases with the number of stars observed simultaneously,
we selected fields that contain a large number of stars 
with V-band magnitudes ($m_{\rm V}$) brighter than 14.0 mag.
The selected fields are thus close to the Galactic plane.
On the other hand, 
most of the Kuiper belt objects are expected to be located near the ecliptic \citep{elliot05}. 
Thus we searched for regions that lie within $2 \degree $ of the ecliptic.
%Finally we have selected the monitoring observation field, 
%whose equatorial and ecliptic coordinates are 
These criteria gave a monitoring observation field with 
equatorial and ecliptic coordinates of
$({\rm RA, Dec}) = (18:30:00, -22:30:00)$ and $({\rm \lambda, \beta}) \sim (276 \degree .9, +0 \degree .8)$, respectively.
%(276 \degree .9267, +0 \degree .75913)

Images of the selected field are obtained simultaneously with the two observation systems
%Images of the monitoring observation field are simultaneously obtained with the two observation systems
for a $2 \times 2$ binned sequential shooting mode of 15.4 frames every second. 
%in $2 \times 2$ binned mode.
The exposure time is $65.0 \ {\rm milliseconds}$ for each frame.
Eight 16-bit uncompressed SER video files (see \url{http://www.grischa-hahn.homepage.t-online.de/astro/ser/}) 
each consisting of 3300 images are produced as a "SER data group" consisting of 26400 images by one capture procedure.
%Between  3300

\begin{figure}[!pt]
\begin{center}
   \includegraphics[scale=0.52]{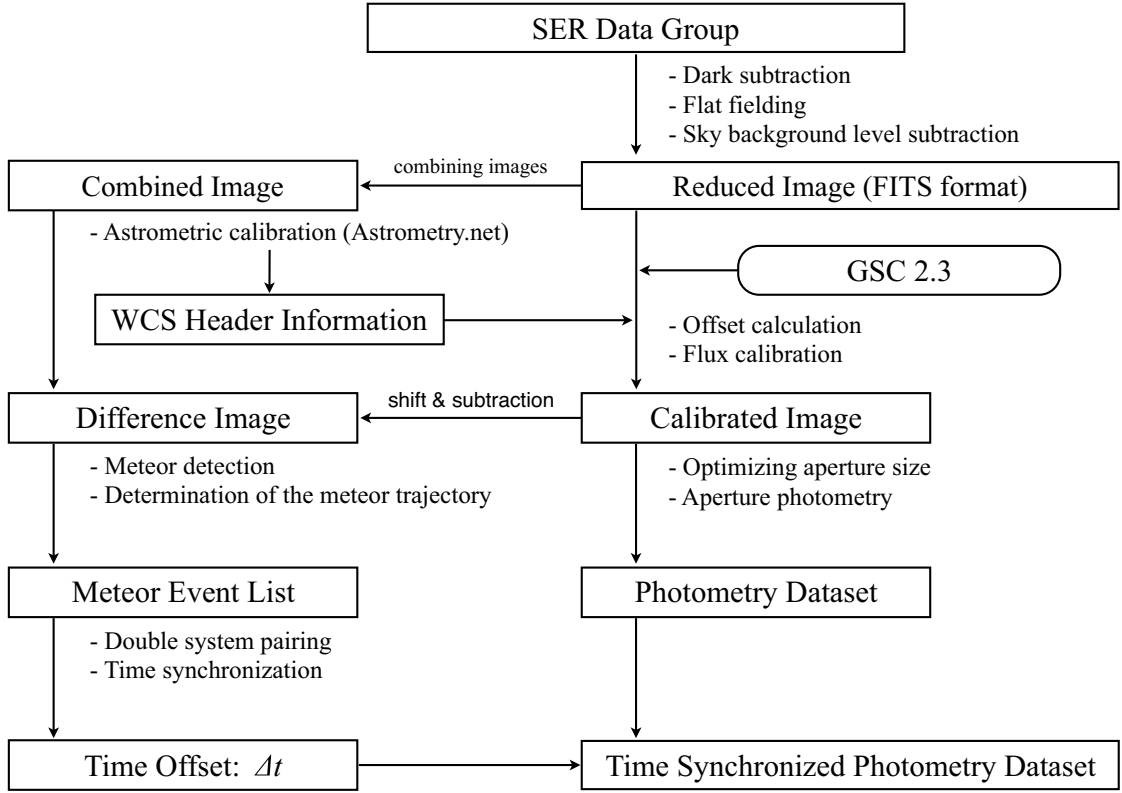}
\caption{Functional overview of the OASES data reduction pipeline. }
   \label{fig43}
 \end{center}
\end{figure}

\begin{figure}[!pt]
\begin{center}
   \includegraphics[scale=0.65]{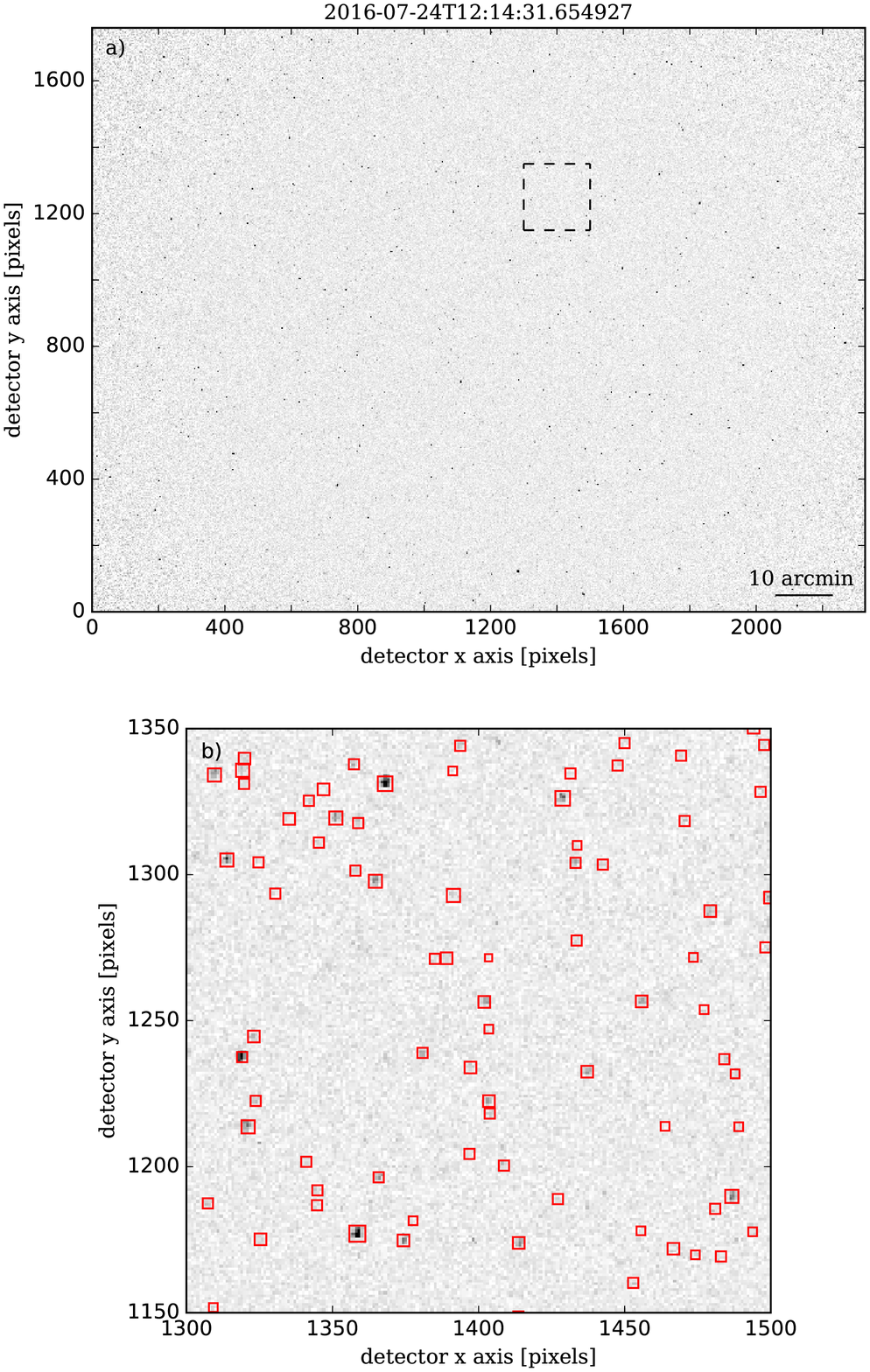}
   \caption{Example of an image obtained in the performance test observations.
   (a) The one-frame reduced image obtained after the dark-subtraction, the flat-fielding, 
   and background subtraction procedures. 
   (b) Enlargement of the image (dashed square in panel (a)), 
   overlaid with the optimal aperture masks for individual stars (red open squares).}
   \label{fig42}
 \end{center}
\end{figure}

\subsection{Data reduction}

%For data reduction of the monitoring observations, 
%we developed the OASES prototypic data reduction pipeline, 
%a Python-based image processing program.
We developed the OASES prototype data reduction pipeline, 
a Python-based image processing program, for data reduction of the monitoring observations.
A functional overview of the pipeline
is shown in figure~\ref{fig43}.
The reduction methods adopted for the pipeline  
are based on the TAOS data reduction procedures \citep{zhang09}
and are adapted for the OASES observations.
The first stage of the pipeline includes dark subtraction using dark frames, 
flat-fielding using twilight flat frames, and subtraction of a constant sky background level, 
to produce FITS-formatted reduced images.
An example of the reduced image obtained 
after these procedures is shown in figure~\ref{fig42}a.
The full-width half maximum of stars in the images is typically $9 - 11\arcsec$,
which corresponds to $\sim$ six physical pixels and $\sim$ three image pixels.

%After the dark-subtraction and the flat-fielding, 
%The first 100 reduced images of the fourth 3300-frame dataset 
A hundred reduced images in every SER data group (the 9901st to 10000th out of the total of 26400 images)
are selected as a sample of the data and are averaged to create a combined image.
%An astrometric calibration for the combined  image is performed using a calibration software 
%{\it Astrometry.net} version 0.67 \citep{lang10}.
An astrometric calibration for the combined image is performed using the {\it Astrometry.net} version 0.67 calibration software 
\citep{lang10}.
Using the World Coordinate System (WCS) parameters obtained after the astrometric calibration,
the pixel coordinates of the stars in the combined image are derived 
with the Guide Star Catalogue (GSC) version 2.3 \citep{lasker08}.
In each frame, pixel coordinates of the stars usually 
change with time due to tracking imperfections of the equatorial mount.
Therefore we derive barycenters of 30 unsaturated bright stars in each frame
to calculate an offset of the pixel coordinates on each epoch. 
%using a centroid-finding algorithm developed in the TAOS studies \citep{zhang09}.

With the pixel coordinates of the GSC stars for each frame, 
we perform aperture photometry for the individual stars
with $m_{\rm V}$ brighter than 14.0 mag. 
Square apertures with sizes $A$ are chosen as photometry apertures
for computational efficiency, 
and a square annulus with a width of $g$ times of the aperture size
is selected as a local background region for each star.
The median value of the pixels in the local background region 
is subtracted from the aperture pixels as a background level
before integrating to produce a signal value. 
The sizes of the square aperture $A$ and the background annulus parameter $g$ for each star
are chosen to optimize the signal-to-noise performance of the resulting light curves.
The aperture-size optimization procedure is 
based on the data reduction method for TAOS \citep{zhang09}.
We perform aperture photometry for each star 
in the 200 (9901st to 10100th) reduced images  in each SER data group
with different aperture and background annulus sizes.
Then we obtain a light curve of 200 data points for each photometry parameter set $(A, g)$.
%in steps of 0.5 pixels from minimum to maximum 
%aperture sizes on the 200 frames of data in eight 3300-frame runs.
After that, we calculate the average intensity $\bar{I}$ 
and the standard deviation of the intensity $\sigma (I) $ for each light curve
to derive the signal-to-noise ratio for the light curve (hereafter {\it light curve S/N}), 
$\bar{I}/\sigma (I)$.
We analyze the light curve S/N
as a function of $A$ and $g$ for each star.
%from a minimum of 2 pixel to a maximum 10.5 pixels  
%in steps of 0.5 pixels for each star.
The minimum aperture size $A_{\rm min}$ of this calculation is chosen to be 2 pixels.
The maximum aperture size $A_{\rm max}$ is determined by the following criteria
in order to minimize contamination from neighboring stars.
%In the present reduction pipeline, 
%a neighboring star $i$ with its V-band magnitude $m_{\rm V} (i)$,
%$m_{\rm V} (i) < m_{{\rm V} t} + 1.5$.
When performing photometry of a star of $m_{\rm V}$ at a pixel coordinate $(X, Y)$, 
we consider neighboring stars with V-band magnitudes brighter than ($m_{\rm V}$ + 1.5) mag. 
%are into consideration.
For any neighboring star $i$ located at $(x_i,y_i)$,
we select the maximum square size $A_{\rm max}$ 
%satisfying 
smaller than the maximum value of two absolute values, $|2(X - x_i)|$ and $|2(Y - y_i)|$.
In the present data reduction pipeline,
$A_{\rm max}$ is chosen in steps of 0.5 pixels from 3.5 to 8 pixels.
If any contaminating star is located within a 3.5-pixel square aperture, 
we did not perform photometry of the target star 
(corresponding to $\sim 6\%$ of the total field stars in the selected field). 

For each star, we perform square aperture photometry 
with aperture sizes $A$ from 2.0 to  $A_{\rm max}$ pixels in steps of 0.5 pixels
and background annulus parameter
$g$ from  $0.5$ to $1.0$ in steps of 0.25.
%is selected to maximize the light curve S/N.
In the present observations, 
about 80\% of the stars have one or more local maxima of the light curve S/N.
We adopt the aperture size where the light curve S/N reaches the local maximum, 
or the aperture size that gives the minimum of the local S/N maxima, 
as the optimized aperture size to minimize contamination from neighboring stars
and noise.
There are cases when the light curve S/Ns increase or decrease monotonically.
In these cases, we adopt the maximum or minimum sizes
as the optimized aperture  sizes, respectively. 
Figure~\ref{fig42}b shows an example of the selected apertures for individual 
stars overlaid on the reduced image.

\subsection{Photometry results}
\label{subsec:photresult}
%The example of the obtained image is shown in 
%the Figure~\ref{fig}. 

\begin{figure}[!pt]
\begin{center}
   \includegraphics[scale=0.9]{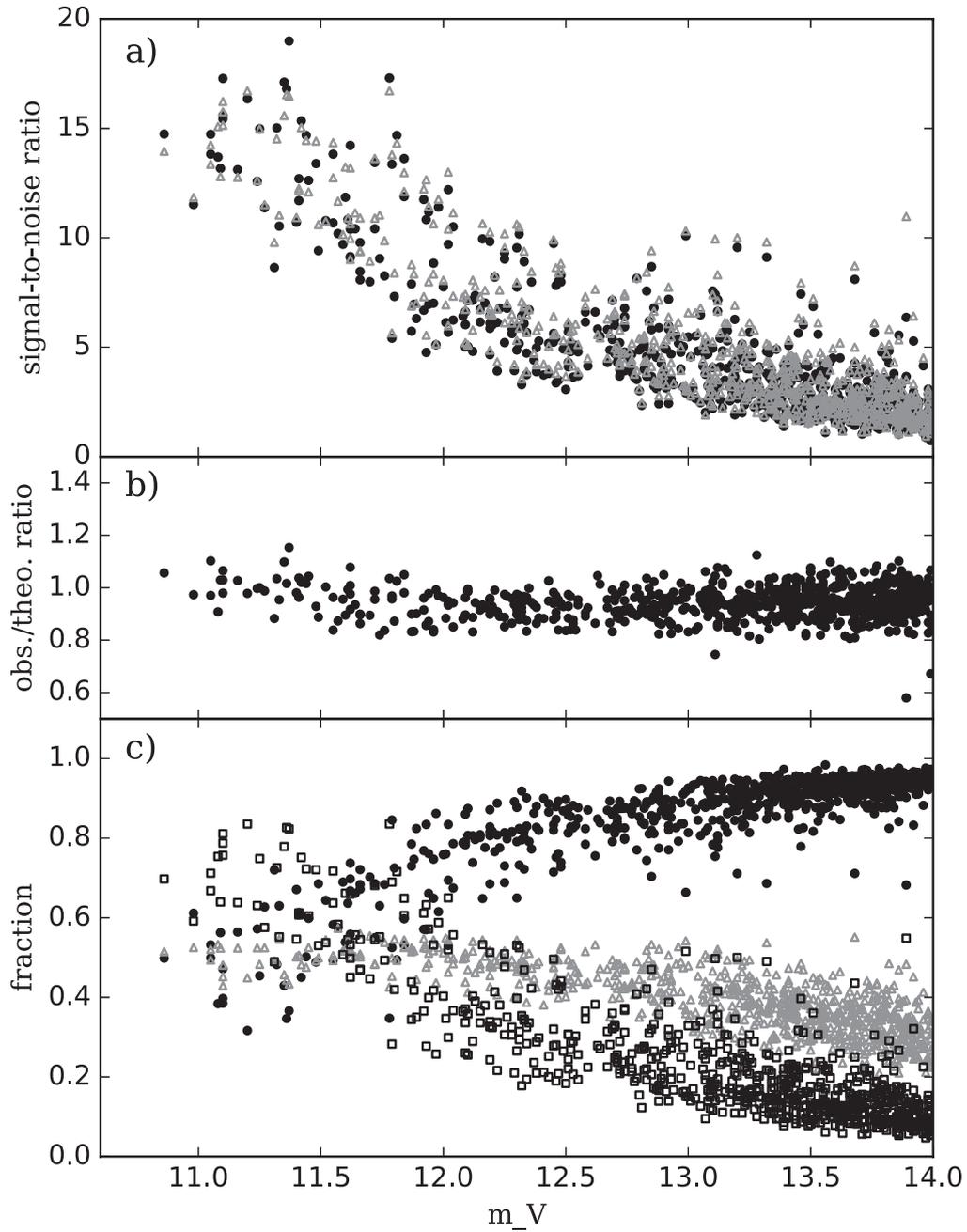}
   \caption{(a) Example of light curve S/N values obtained 
   in the OASES monitoring observations on a dark night (filled black circles) and theoretical S/Ns 
   (open grey triangles) 
   as a function of $m_{\rm V}$.  
   Only a sample of 8\% of all stars is shown for clarity.
   In this plot, $\eta$ is assumed to be $0.05$. 
 (b) Observed light curve S/Ns scaled by the  theoretical light curve S/Ns.
 (c) The variance of major noise sources; the readout noise ($\sigma_{\rm read}$; filled black circles),
 the target shot noise ($\sigma_{\rm source}$; open grey triangles), and the scintillation noise ($\sigma_{\rm sci}$; open black squares),
 scaled by the total theoretical noise.}
   \label{fig4r3}
 \end{center}
\end{figure}

Figure~\ref{fig4r3}a shows an example of the light curve S/Ns 
obtained from the OASES monitoring observations (plotted as black filled circles).
%These values are evaluated from the r.m.s. values of the light curves 
%for individual stars divided by the average signal values. 
The average S/N value reaches 3, 5, and 10 at $m_{\rm V} \sim 13.5$, $12.8$, and $11.8$, respectively.
%These values 
%The SN reaches an constant value ($\sim 13$) at  $m_v < 10$.
%In order to examine the photometric performance of the OASES 
%with the measured light curve S/N values, 
In order to characterize the noise performance of the OASES observation system, 
we compare the observed light curve S/Ns 
with the theoretical S/N values for the individual stars.
In general, the light curve S/N for a star  
obtained with high-cadence observations
using an optical photoelectric detector array 
would be approximated by the following equation: 
\begin{eqnarray}
\textcolor{red}{
%\label{eq_phot}
	%{\rm theoretical} \ S/N = \frac{\kappa \cdot S_{star}}{\sqrt{\sigma_{read}^{2}  + \sigma_{dark}^{2} + \sigma_{sky}^{2} + \sigma_{source}^{2} +\sigma_{sci}^{2}}}
	{\rm theoretical} \ S/N = \frac{S_{\rm star}}{\sqrt{\sigma_{\rm read}^{2}  + \sigma_{\rm dark}^{2} + \sigma_{\rm sky}^{2} + \sigma_{\rm source}^{2} +\sigma_{\rm sci}^{2}}}
	}
\end{eqnarray}
where 
%$\kappa$ is the conversion gain factor in $e^{-}$ ADU$^{-1}$, 
$S_{star}$ is the signal values of a star obtained with the square aperture size  $A$ in $e^{-}$. %ADU.
$\sigma_{\rm read}$, $\sigma_{\rm dark}$, $\sigma_{\rm sky}$, $\sigma_{\rm source}$, and $\sigma_{\rm sci}$
represent the detector readout noise, the dark noise, the sky background noise, the target shot noise, and the scintillation noise, respectively, 
which are expressed as follows:
\begin{eqnarray}
\sigma_{\rm read}^2 &=& \sigma_{\rm read_{\rm pix}}^{2}  A^2, \\
\sigma_{\rm dark}^2 &=& i_{\rm dark}   A^2,  \\
\sigma_{\rm sky}^2     &=& s_{\rm sky}  A^2,  \\ %A^2 \, (1+1/(mB^2-1)) \\
\sigma_{\rm source}^2 &=& S_{\rm star},   \\
\sigma_{\rm sci}^2 &=&  \eta^2 S_{star}^2,
\end{eqnarray}
where 
$\sigma_{\rm read_{\rm pix}}$ is the standard deviation of the detector readout noise in $e^{-}$ pix$^{-1}$,
$i_{\rm dark}$ is the dark current for the observed exposure time (65 milliseconds) in $e^{-}$ pix$^{-1}$,
$s_{\rm sky}$ is the surface brightness of the sky background in $e^{-}$ pix$^{-1}$,
and $ \eta $ is the fractional scintillation variance relative to $S_{star}$.
%$n'_{\rm pix} $ is the effective number of pixels for the aperture photometry.
$A^2$ corresponds to the effective number of pixels for the square aperture photometry size $A$. 

According to the laboratory tests, 
$\sigma_{\rm read_{\rm pix}}$ is estimated to be $5.8  \  e^{-}$ pix$^{-1}$
in the $2 \times 2$ binned mode.
On the other hand, $i_{\rm dark}$ is estimated to be smaller than $0.1  \  e^{-}$ pix$^{-1}$ 
at the operating sensor temperature ($5-10 \ {}^\circ\mathrm{C}$).
The contribution of $\sigma_{\rm dark}^2$ relative to $\sigma_{\rm read}^2$ can thus be considered negligible. 
\textcolor{red}{
$s_{\rm sky}$ is determined from the 3-$\sigma$ clipped average pixel values 
of images obtained in the observations. 
%measured with the image 
The measured $s_{\rm sky}$ is $5-10 \  e^{-}$ pix$^{-1}$,
}
and thus $\sigma_{\rm sky}^2$ is insignificant relative to the $\sigma_{\rm read}^2$.

The theoretical light curve S/Ns for the individual observed stars are 
plotted as open triangles in figure~\ref{fig4r3}a 
and are compared with the observed S/Ns in figure~\ref{fig4r3}b. 
In figure~\ref{fig4r3}, $\eta$ is approximated as $0.05$.
%The  theoretical S/Ns well reproduce the observed light curve S/Ns 
%in the observed flux ranges.
According to the comparison between the theoretical S/Ns and the observed values, 
the ratio of observation to theoretical S/N approaches unity
when the fractional scintillation variance $\eta$ is taken as $0.05-0.1$.
This fractional variance is consistent with that obtained in the performance test 
observations of bright stars ($\sim 0.05-0.08$, see subsection~\ref{subsec:test}).
With this $\eta$ value,
the observed light curve S/Ns are well approximated by the theoretical performance.
Figure~\ref{fig4r3}c shows the variance of each major noise source ($\sigma_{\rm read}$, $\sigma_{\rm source}$, and $\sigma_{\rm sci}$) 
scaled by the total theoretical noise. 
At the flux range of $m_{\rm V} \sim 12-14$, 
the detector readout noise is the dominant noise source.
On the other hand, at the brighter flux range ($m_{\rm V} < 11.5$),
the  target shot noise and the scintillation noise become 
comparable with the  readout noise.
%One can therefore consider the typical noise source in the OASES data  
%is the combination between the readout noise 
%and the shot noise from the source.

\begin{figure}[!pt]
\begin{center}
   \includegraphics[scale=1]{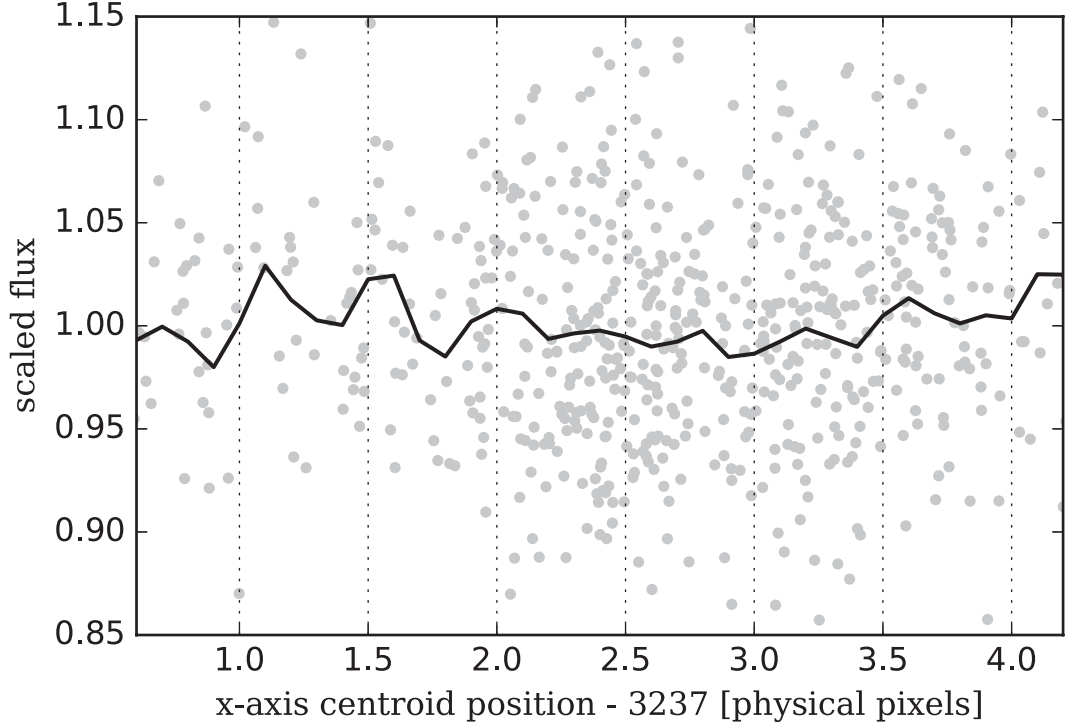}
   \caption{
   \textcolor{red}{
   Example of a scatter plot showing the measured fluxes of a bright star (TYC 6857-238-1, $m_{\rm V} = 10.7$) 
   scaled by their average value against the centroid positions in the x-axis direction of the CMOS sensor.
   In this example, the star moves in the hour angle direction 
   (parallel to the x-axis direction of the CMOS sensor) in a two-minute observation
   due to tracking imperfections of the mount.
   A solid line represents moving average of the fluxes with a window of 0.1 pixels. 
   The standard deviation of the fluxes and of their moving average values
   are 0.058 and 0.014, respectively.
   For the data points used in the plot, 
   the relative differences of the positions in the y-axis direction  are smaller than 0.2 pixels.
   }
   }
   \label{fig_cent}
 \end{center}
\end{figure}

\textcolor{red}{
The measured stellar flux may be affected by non-uniformity of sensitivity within the pixel,
because the OASES observation system uses a commercial front-illuminated CMOS sensor,
in which readout electronics decreases the photon collecting area within the pixel. 
It can cause a systematic error on the flux with imperfect tracking and degrade the detectability 
of stellar occultations. 
In order to examine how the non-uniformity within the pixel affects the stellar flux measurements, 
we carried out a comparison between the measured flux values of a bright star 
and their centroid positions obtained in the monitoring observations with imperfect tracking. 
An example of a scatter plot of the measured fluxes against the centroid positions of a bright 
is shown in Figure~\ref{fig_cent}. 
This plot shows no clear trend of the flux with an amplitude greater than $\sim 2\%$ of the average flux values. 
The systematic uncertainty is thus  significantly small compared to the depth of the stellar occultations by km-sized 
TNOs (typically larger than  $20\%$) and is expected to be negligible for the current occultation observations.
%Since the OASES observation system uses a commercial used front-illuminated CMOS sensor,
%not a scientific grade back-illuminated sensor, 
%the measured stellar flux may depend on the pixel coordinate of the star
%due to non-uniformity of sensitivity within the pixel.
%This non-uniformity is expected to be worse for the front-illuminated CMOS sensor
%because readout electronics decreases the photon collecting area within the pixel.
%Even though the non-uniformity effect is apparently small and is not clearly evident 
%in the observed light curve S/Ns, 
%it can cause a systematic error on the flux obtained by observations with imperfect tracking 
%and degrade the detectability of stellar occultations.
%%imperfect tracking of the mount during the observations 
%%may cause systematic effects on the measured flux of stars.
%%Due to the non-uniformity of the sensitivity within the pixel of the sensor,
%%the flux may depend on centroid position of the star.
%%In general, oversampled ($>$)
}

%\textcolor{red}{
%In order to examine how the non-uniformity within the pixel affects the stellar flux measurements,
%we carried out a comparison between the measured signal values of a bright star 
%and their centroid positions obtained in the monitoring observations with imperfect tracking.
%An example of a scatter plot of the measured signals vs the centroid positions of a bright star 
%obtained in the monitoring observations is shown in Figure~\ref{fig_cent}.
%The measured signal values and their moving average values (gray circles and a solid line in Figure~\ref{fig_cent}, respectively) 
%do not show clear variations, 
%which are expected to be correlated with the pixel scale 
%if the non-uniformity effect is present.
%Due to the limited precision of the measured flux values,
%we do not rule out the possibility of the non-uniformity effect 
%with an amplitude of the variations smaller than $\sim 2 \%$. 
%However, the possible amplitude is significantly small
%compared to that of the stellar occultations 
%by km-sized TNOs (typically larger than $\sim 20\%$) 
%and is expected to be negligible for the current occultation observations.
%}

\begin{figure}[!pt]
\begin{center}
   \includegraphics[scale=1]{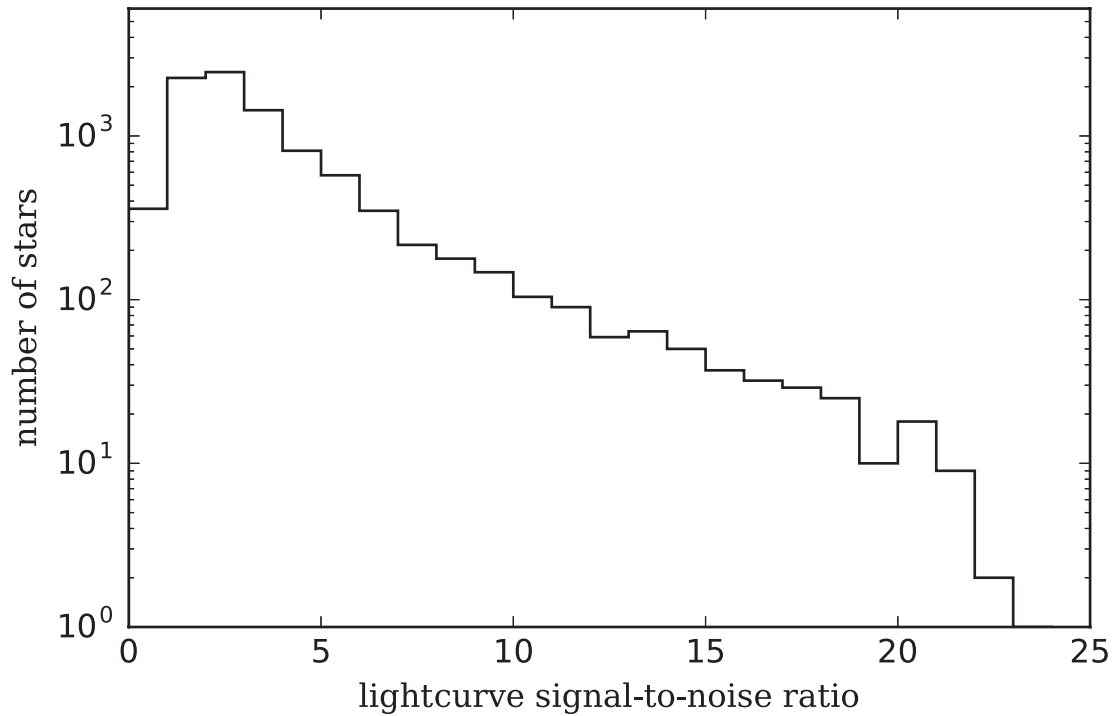}
   \caption{Histogram of the number of stars 
   observed simultaneously with the OASES observation system as a function of the light curve S/N.}
   \label{fig5}
 \end{center}
\end{figure}

Figure~\ref{fig5} shows the histograms of the number of stars simultaneously
observed with the OASES observation system as a function of the light curve S/N.
For the selected observation field, 
the observation system is capable of monitoring $\sim 500$, $2000$, and $4200$ stars 
with light curve S/Ns greater than 10, 5, 3, respectively.
According to \citet{nihei07},
the detectability of the occultation depends highly on the TNO size,
the sampling rate of the observation, and the light curve S/N. 
Assuming a stellar occultation by a Kuiper-belt ($R_h = 40$ au) TNO at opposition
observed with the two OASES observation systems at a sampling rate of 15.4 Hz,
an occultation event by a TNO with a 
%diameter $D$ of greater than $\sim 2$, $3$, and $6$ km 
\textcolor{red}{ radius $r$ of greater than $\sim 1$, $1.5$, and $3$ km}
can be detected from light curves with the light curve S/N of 10 and 5, and 3, 
respectively, with the false positive rate lower than $10^{-11}$.
Therefore stellar occultations by km-sized TNOs are detectable with the light curves.
In the previous coordinated observations (TAOS), 
the number of stars that could be monitored simultaneously 
was up to $\sim 1000$ \citep{lehner10} with a sampling rate of $5 \, {\rm Hz}$.
%and the typical S/N of the stars are $\sim 3-4$  \citep{bianco10}. 
Therefore the OASES observation system is capable of monitoring 
a larger number of stars simultaneously with the greater sampling cadence
and will provide the statistically useful amount of photometry data 
needed to improve the size information of km-sized TNOs.

%The obtained results indicate that the OASES observation system is capable of monitoring 
%a larger number of stars simultaneously with the greater sampling cadence
%and will provide the statistically useful amount of photometry dataset 
%needed to improve the size information of the km-sized TNOs.

\section{Time synchronization using faint meteors}
\label{meteor}

\begin{figure}[!pt]
\begin{center}
   \includegraphics[scale=0.75]{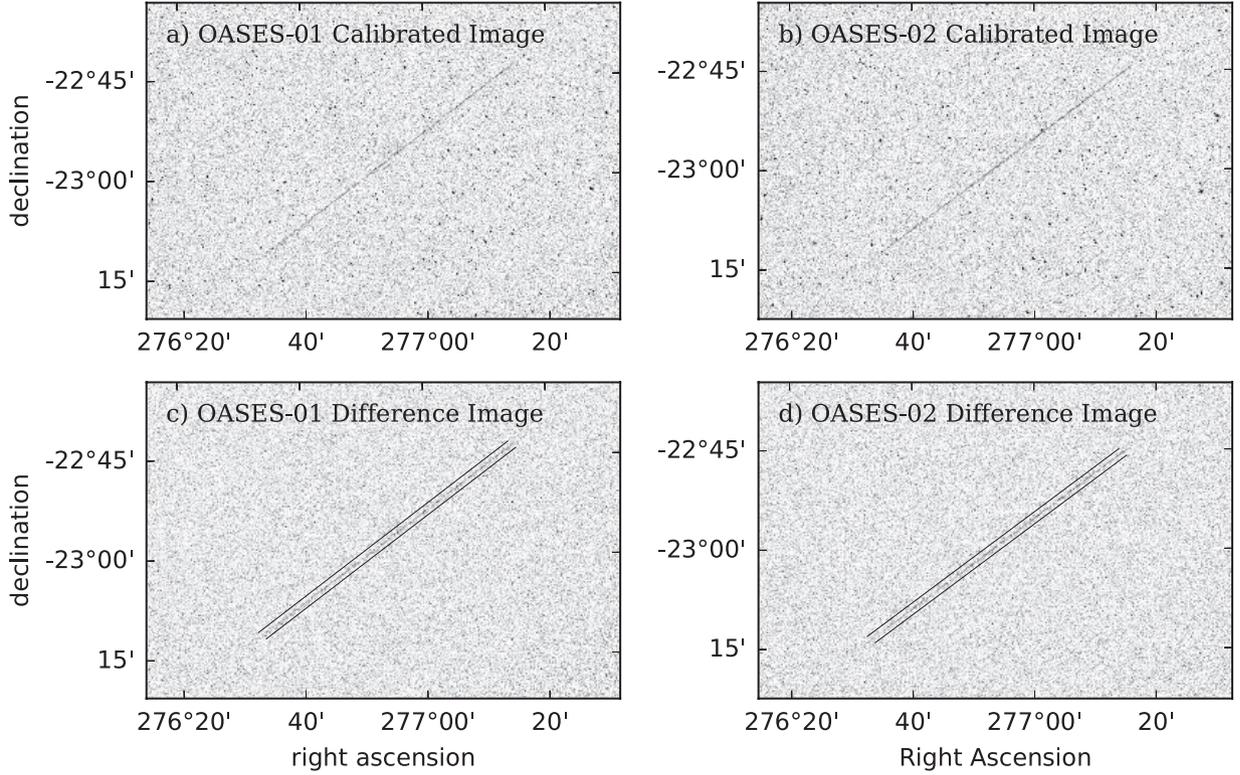}
   \caption{Example of a faint meteor detected by the OASES monitoring observations.
   Panels (a) and (b) are calibrated images of the faint meteors 
   %(with an integrated brightness of $9.0$ mag) 
   obtained with OASES-01 and OASES-02, respectively.
   Panels (c) and (d) are the same as (a) and (b), but after the subtraction and masking of stationary features.
   The faint meteor is highlighted by two parallel line segments.
  }
   \label{fig44}
 \end{center}
\end{figure}

Precise time synchronization between the observation systems
is essential for the coordinated occultation observations.
%In order to detect stellar occultations simultaneously,
%A precise time synchronization between observation systems is required
%for the OASES observations.
The internal clock of the OASES control PC is thus maintained by 
the time information obtained with the USB GPS receiver, as described in section~\ref{sec:obssystem}.
However, the GPS time synchronization did not work perfectly in the 2016-season monitoring observations.
We found that the timing of the appearances of meteors that serendipitously appeared 
in the observation data (see figure~\ref{fig44})
is inconsistent between the two observation systems. 
%of the SER header files between the two observation systems
%by   
%
The timestamps in the SER header files contain an offset from the reference time.
%which is confirmed to be constant during the operation of the control PC, 
This time offset appears to be constant (typically $1-5$ seconds) during the turn-on of the control PC 
and varies randomly with observation date.
%\textcolor{red}{
%and is typically $1-5$ seconds.
%}
The origin of the offset is thought to be imperfect time synchronization 
between the image capture software and the operating system of the control PC.
Therefore an alternative calibration was needed for the time synchronization.

For the current dataset, we carried out an additional timing calibration 
using meteors 
%serendipitously 
that appeared in the observation data (figure~\ref{fig44}). 
We developed a detection algorithm based on the 
meteor line detection method using the Hough Transform \citep{gural08}.
This Python-based  algorithm finds bright lines (or meteor trajectories)
in a calibrated image after the subtraction and masking of stationary features such as stars 
(difference image; figure~\ref{fig44}c and d, see also figure~\ref{fig43}).
The current detection algorithm is capable of detecting meteors 
with integrated brightness as faint as $\sim 10$ mag.
%which are emerged roughly $\sim 30$ times per one night observation run.
For the timing calibration, we select meteors detected in more than two frames.
With this criterion, 
the selected meteor should be detected as a short "line segment" in one or more frames. 
We thus measure the coordinates of the starting and end points of each line segment
to derive the celestial coordinates of the meteor as a function of the uncalibrated time (derived from the timestamp), 
assuming a uniform linear motion. 
If there are two or more trajectories obtained in individual frames, 
we use the trajectory whose central position is the closest to the center of the image in the present calibration.
If the integrated fluxes of the trajectories show a large variation, 
we use a trajectory whose integrated flux is the maximum for the calibration.

The time offset value between the two observation systems $\Delta t$  
is derived by minimizing ${\rm d}a$ defined by the following function;
\footnotesize
\begin{eqnarray}\label{eq_da}
%\begin{split}
%da = \frac{1}{N} \sum_{k = 1}^{N}  \arccos(\sin{\delta_{1k}} \sin{\delta_{2}(t_{1}+\Delta t)} + \cos{\delta_{1k}} \cos{\delta_{2}(t_{1}+\Delta t)\cos(\alpha_{1k} - \alpha_{2}(t_{1}+\Delta t))}),
{\rm d}a = \arccos(\sin{\delta_{1}(t_{1})} \sin{\delta_{2}(t_{1}+\Delta t)} + \cos{\delta_{1}(t_{1})} \cos{\delta_{2}(t_{1}+\Delta t)\cos(\alpha_{1}(t_{1}) - \alpha_{2}(t_{1}+\Delta t))}),
%\end{split}
\end{eqnarray}
\normalsize
where $\alpha_i (t)$ and $\delta_i (t)$ 
are the right ascension and declination of the meteor observed with an observation system OASES-$i$ 
at an uncalibrated time $t$.
%These values are derived from the linear approximation using  in the image obtained  with the OASES-$i$ system.
%($\alpha_{i}$, $\delta_{i}$) is the central position of the line segment, and $t_i$ 
%is an uncalibrated timestamp of the frame.  
${\rm d}a$ thus corresponds to the average angular distance of the same meteor 
observed with the two OASES observation systems (OASES-01 and OASES-02) 
at the uncalibrated time of $t_1$ and $t_1 + \Delta t$, respectively. 
Assuming that the parallax of the meteors can be ignored
\footnote[1]{The parallax of a meteor observed from two systems is expected to be $\sim 1\arcmin.3$ at most, 
assuming the altitude of the meteor is $100\ {\rm km}$.
Since the typical angular velocity of the detected meteors is $\sim 5\degree \ {\rm sec^{-1}}$,
the possible timing error due to the parallax is $\sim 5 \ {\rm milliseconds}$ at most.},
the true offset $\Delta t$ offers the minimum value of ${\rm d}a$.
We calculate $\Delta t$ for the individual meteors detected in two observation systems
and adopt the average value as the offset for the time calibration.

\begin{figure}[!pt]
\begin{center}
   \includegraphics[scale=0.85]{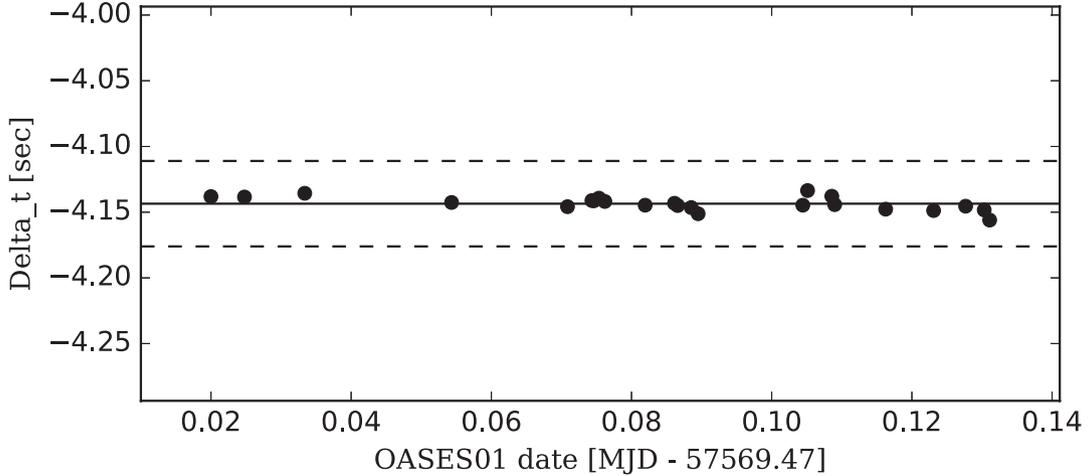}
   \caption{Example of the time offset value $\Delta t$ obtained 
   with the time synchronization method using meteors.
   Filled circles represent the $\Delta t$ values derived from the individual meteors.
   The solid line represents the average $\Delta t$ value.
   Dashed lines correspond to the one-frame exposure time (65 millisecond),
   which is shown for reference.
   In this example, the constant offset value used for the time calibration 
   corresponds to $-4.144$ seconds with a $1\sigma$ uncertainty of $0.005$ seconds. 
   %The offset value corresponds to half of the exposure time for each frame (65 millisecond).
  }
   \label{fig45}
 \end{center}
\end{figure}

%This calibration method has been used for time synchronization of 
%10 data runs comprising a total of 5 hour observation data
%in order to verify the the validity of the technique. 
Figure~\ref{fig45} shows an example of the time synchronization results.
The detection rate of the meteors for the calibration 
is roughly $\sim 5$ to $10$ times per one-hour observation run.
After the calibration with the constant offset value, 
the timing between the two systems 
is synchronized with the $1\sigma$ accuracy of $\sim 5$ milliseconds, 
which is comparable to 1/10 of the one-frame exposure time
and is sufficient for the present study.

%%%%%%%%%%%%%%%%%%%%%%%%%%%%%%%%%%%%%%%%%%%%%%%%%%%%
\section{Summary and future prospects}
%%%%%%%%%%%%%%%%%%%%%%%%%%%%%%%%%%%%%%%%%%%%%%%%%%%%

We have presented an optical, high-speed, wide-field
observation project, OASES,
that aims to investigate stellar occultation events by km-sized TNOs.
Two low-cost observation systems 
consisting of commercial off-the-shelf 0.28 m aperture f/1.58 optics 
and a CMOS camera  were developed in OASES.
%Stellar occultation events by the TNOs can be detected 
%from the intensity variations of the stellar light curves
%that will be obtained with the combinations of latest mass-produced compact telescopes and CMOS detectors.
The results of the monitoring observations at Miyako Islands, Okinawa, performed in 2016
demonstrate that the OASES systems are capable of high-speed ($15.4 \ {\rm Hz}$ cadence) 
simultaneous photometry of $\sim 2000$ stars with magnitudes down to $m_{\rm V} \sim 13$, 
providing $\sim 20\%$ photometric precision in the light curve. 
%The characterization of the noise on the obtained light curves shows 
%that it is dominated by the contribution of the known readout noise
%and of  the object photon noise (brighter objects). 
We also developed a time synchronization method 
using faint meteors, which is needed for the 
simultaneous occultation detection.
These developments enable us to achieve robust 
detection of stellar occultation events 
with the two independent instruments simultaneously.

This paper describes the present OASES observations for monitoring
stellar occultations by TNOs.
%This paper describes the present OASES monitoring observations 
%of stellar occultations by the TNOs.
However, the OASES observation systems 
are capable of exploring other rare and short-timescale astronomical events,
such as faint meteors, lunar impact flashes,  near-Earth asteroids and space debris.
%In fact, dozens of faint meteors detected in one night's OASES observations simultaneously
In fact, dozens of faint meteors are detected in one night of OASES observations simultaneously
(see section~\ref{meteor}).
Since these astronomical events are rare and non-repeatable,
simultaneous detection with multiple instruments is essential 
for their statistical analysis.
Thanks to the unique capability of wide-field high-cadence imaging 
and the cost-effectiveness of the OASES observation systems, 
OASES will be key instruments for pioneering observations 
in high time resolution astronomy.

As of 2017, the OASES project has developed two observation systems.
\textcolor{red}{
A primary monitoring observation campaign with the two systems began in June 2016
and is now underway.
We plan to continue the observation campaign with the two systems until 2018.
}
%However, unlike other observation programs, 
%one can easily increase the number of the observation systems because of its low cost.
%As noted in the previous studies \citep{lehner10},
%adding the systems significantly reduces the false-positive detection rate.
%Therefore the OASES observations would 
%be capable of achieving more robust detections of the occultations as additional systems are installed.
%We also note that the instruments used for the present OASES observation system
%are easily available for amateur astronomers.
%The present observation project thus demonstrates
%an opportunity for them to participate in cutting-edge astronomical studies.
However, unlike other observation programs, 
the number of observation systems can be easily increased because of their low cost.
As noted in previous studies \citep{lehner10},
adding systems significantly reduces the false-positive detection rate.
Therefore the OASES observations will 
be capable of achieving more robust detection of occultations when additional systems are installed.
\textcolor{red}{
%This future upgrade is currently under consideration.
%The schedule of adding the systems is currently being developed.
This upgrade is currently in its conceptual design phase.
}
We also note that the instruments used for the present OASES observation system
are easily available for amateur astronomers.
The present observation project thus demonstrates
an opportunity for them to participate in cutting-edge astronomical studies.

\bigskip
%We gratefully acknowledge the entire contributions of Kiso observatory.
We thank the personnels of Miyakojima City Museum
and of Miyako open-air school %({\it Miyako Shonen Shizen no Ie}) 
\textcolor{red}{
({\it Miyako Seishonen no Ie})
}
for cooperating 
on the OASES observations at Miyako Island.
We also thank the people of the in Miyako Island for supporting our observations.
%especially Mr. Keiji Nakasone, Ms. Kaori Ikeshiro, and Ms. Fumika Yonaha. 
%PyRAF is a product of the Space Telescope Science Institute, which is operated by AURA for NASA.
This research has been partly supported by JSPS grants (JP26247074, 15J10278, 
\textcolor{red}{
26800112,
}
and 16K17796).
%The authors are grateful to Mr.  for advice on proofreading the English manuscript.

\end{document}